# A Federated Learning Framework for Non-Intrusive Load Monitoring

Haijin Wang[1] *Student Member*, IEEE, Caomingzhe Si[1] *Student Member*, IEEE, Junhua Zhao[1] *Senior Member*, IEEE

[1]School of Science and Engineering, The Chinese University of Hong Kong, Shenzhen, CHN

*Abstract*- Non-intrusive load monitoring (NILM) aims at decomposing the total reading of the household power consumption into appliance-wise ones, which is beneficial for consumer behavior analysis as well as energy conservation. NILM based on deep learning has been a focus of research. To train a better neural network, it is necessary for the network to be fed with massive data containing various appliances and reflecting consumer behavior habits. Therefore, data cooperation among utilities and DNOs (distributed network operators) who own the NILM data has been increasingly significant. During the cooperation, however, risks of consumer privacy leakage and losses of data control rights arise. To deal with the problems above, a framework to improve the performance of NILM with federated learning (FL) has been set up. In the framework, model weights instead of the local data are shared among utilities. The global model is generated by weighted averaging the locally-trained model weights to gather the locally-trained model information. Optimal model selection help choose the model which adapts to the data from different domains best. Experiments show that this proposal improves the performance of local NILM runners. The performance of this framework is close to that of the centrally-trained model obtained by the convergent data without privacy protection.

## I. Introduction

ENERGY disaggregation, also known as NILM, monitors various types of appliance-level power consumption through the disaggregation of total electric load, instead of installing sensors on the entries of appliances. NILM measures and analyzes the total electric load of households by installing advanced metering infrastructures (AMI) only at the electrical entry to measure signals such as voltage, current, active, and reactive power, etc. By analyzing the appliance-wise information, the total household power consumption would decrease by 5%-20% [1]. From the perspective of economic installation, AMI only records the whole-house information in current practice, and promotes the development of NILM [2], for it is a cost-effective alternative other than installing distributed smart meters on appliances. To this end, NILM disaggregates the overall consumption into appliance-wise ones to lay solid foundations for tasks such as energy conservation and emission reduction etc. To better accomplish NILM, plenty of methods has been put forward such as expert heuristics methods that consist of the creation of a set of rules for each appliance [3-6], decision trees, long short-time memory that used for event detection [7, 8]. Methods based on deep learning have become a hotshot in the field of NILM in recent progress.

To train a deep network that maps between the total electric load reading and appliance-wise power consumptions, sequence-to-point learning (seq2point) that employees both convolutional neural networks (CNN) and recurrent neural networks (RNN) overweigh sequence-to-sequence learning [9] in improving the efficiency of NILM [10]. Given the total electric load series as an input, the trained seq2point network engenders the output signals of each appliance at the midpoint of the input time series. Deep neural networks (DNN) such as CNN and RNN [11, 12] in which have numerous intrinsic parameters to be trained [13] are ignited for the abundant training data that accessible in recent years [14-16]. On the other hand, to obtain better generalizations on an extensive range of households and appliances, it is essential to gather massive electric load data from different local runners. In practical, utilities and distributed network operators (DNO) usually play the role of local runners.

Though model performances of DNNs can be improved via cooperation among utilities and DNOs, multi-source NILM data may bring about privacy disputes [17] for there exist chances to acquire personal privacy of households. To be specific, when NILM data is gathered at one coordinating server, the consumers' privacy may be faced with the risk of theft for the high frequency of information exchange. Moreover, utilities and DNOs may lose their data control during data transmission. Local obfuscation [18] is widely used in privacy protection and data control conservation. While in the smart grid scenario, local differential privacy (LDP) that integrates local obfuscation has always been a most welcome solution [19]. Distinct from approaches that extract exact power consumption, LDP adds noise selectively on the original electric load series to provide protection. For tasks that applied NILM, goals are not second-by-second reconstructions of each appliance. On the contrary, for each appliance activation, both the identification of start-stop time and electricity consumed is needed for the most part. Excessive noise may distort the training data in the co-training process, on the other hand, minor noise can provide few privacy protections on the original power consumptions. Therefore, it is hard to achieve a balance between consumer privacy protection and data utilization during the cooperation among utilities and DNOs.

Data cooperation is promoted through the proposed framework taking advantage of FL [20] and the performance of the co-training model is improved during the process. The principle of FL is to share update parameters of locally-trained models among participants without the need for data transfer. There would be several merits applying the proposed framework. Firstly, consumers' privacy is protected as the households' electric load data can never be leaked. Secondly, the proposed FL framework enables utilities and DNOs to manage household electric load data without the expensive and fragile data replication or transmission, thereby releasing the requirement of network bandwidth involved with the update of datasets on a different scale. Thirdly, utilities and DNOs would never worry about the loss of data control as data is stored in each participant's database individually instead of being concentrated on one server, thereby reducing the probability of data exfiltration.

This paper has made the following contributions. Firstly, an efficient FL framework is unprecedentedly adopted to train a neural network for NILM. Both the improvement of model performance and the goal of privacy protection are achieved simultaneously. Secondly, the model obtained by the proposed framework outperforms the average level of locally-trained models and approximates the performance of centrally-trained one, this demonstrates the effectiveness of the proposed framework in co-modelling. Thirdly, as the number of runners increases, the improvement of model validity is verified.

The remainder of the paper is organized as follows. Section 2 gives a detailed summary of prior work related to NILM and FL. Section 3 makes a description on the overview of the framework architecture. Section 4 demonstrates the experiment of proposed framework. Section 5 presents the experiment results and the evaluations. Section 6 gives the concluding remarks.

## II. RELATED WORK

### A. DNN-based Non-Intrusive Load Monitoring

Apart from traditional approaches that have been proposed for NILM, DNN-based methods [22, 23] relies on creating new networks for households and appliances. With the availability of sufficient and quality training datasets, DNN performs well as they are highly targeted. Among all DNN methods that can be effectively utilized in NILM, seq2point learning is a data-driven approach requiring only data and labels.

#### 1) Brief Introduction of Seq2point Learning

Seq2point learning is employed by utilities and DNOs to accomplish NILM tasks for its outstanding performance in NILM [10]. In the scenario of the paper, seq2point learning identifies the continuous time series of the total electric load and decomposes it into a binary discrete time series of multiple appliances' power consumption. Given the window of total electric load $Y_{t:t+W-1}$ as an input, seq2point learning trains a neural network to represent the midpoint of the selected window for each appliance. Specifically, the output is the midpoint element value of the targeted appliance for the corresponding window. The detailed architecture of seq2point learning is shown in Figure 1. In the proposed framework, seq2point learning helps to obtain a non-linear mapping between the time series of total electric load and the midpoint element of targeted appliances. The method assumes that the state at the midpoint element of the targeted appliance is related to the entirety status before and after the midpoint.

#### 2) Architecture of Seq2point Learning

Seq2point learning is applied to find out network $F_p$ which maps sliding windows $Y_{t:t+W-1}$ of the mains to the midpoint value $x_\tau$ of corresponding windows $X_{t:t+W-1}$ of the output. The output of the model is $\widehat{x_\tau}$.

The mapping is modeled as,

$$x_\tau = f(Y_{t:t+W-1}) + \varepsilon \quad (1)$$

The loss function $L_p$ for training takes the form,

$$L_p = \sum_{t=1}^{T=W+1} (x_\tau - \widehat{x_\tau})^2 \quad (2)$$

where $\theta$ are the parameters trained by the network. The advantage of seq2point learning is that there is a single prediction for each $x_t$ other than an average of predictions for each window.

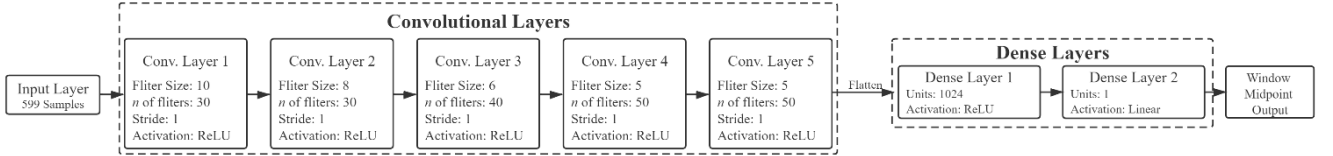

Fig. 1. Architecture for seq2point learning

### B. Federated Learning

FL attends in modelling across multiple decentralized devices or servers holding local samples without the process of data exchange, therefore data privacy, data security, and data exclusivity can be promised. The current applications of FL are industries including defense, telecommunications, internet of things, and pharmaceutics [20]. Till now, it has not been applied in the field of NILM. In 2017, the typical federated averaging (FedAvg) raised by google [21] is applied to help several parties model jointly.

There are two key parts in FedAvg. On the one hand, local runners update iteratively their locally-trained model's parameters using local training data. On the other hand, the coordinating server gets a weighted average of the locally-trained models' parameters. to get the global model. The above process is run for multiple rounds.

$$w_{t+1} \leftarrow w_t - \eta \sum_{k=1}^{K} \frac{n_k}{n} g_{k,t} \quad (3)$$

where $w_{t+1}$ denotes the global model's parameters of round $t+1$, $w_t$ denotes the global model's parameters of round $t$, $\eta$ denotes the learning rate of global parameter updates, $n_k$ and $n$ are the number of total training samples and the number of training samples on local runner $k$, respectively. $g_{k,t}$ is the gradient obtained by local runner $k$ of round $t$.

While in the locals, runners iterate the following local update multiple times before the averaging step.

$$w_{e+1} \leftarrow w_e - \eta \cdot \delta l_e \quad (4)$$

where $w_e$ denotes the locally-trained model's parameters of epoch $e$ during the "Client Update", $\eta$ denotes the learning rate of local parameter updates. $\delta l_e$ denotes the gradient obtained by local runner of epoch $e$ during the "Client Update".

## III. FRAMEWORK ARCHITECTURE

### A. Goal of Federated Learning Non-Intrusive Load Monitoring Framework

The goal of the proposed framework is to promote co-modelling among runners to improve the performance of NILM while protecting consumer privacy and conserving data control. In 2017, the original FedAvg proposed by Google is to help obtain a shared global model for all decentralized local nodes with the consideration of privacy protection. The main idea is averaging the parameters in locally-trained models weighted without accessing users' information. The framework is as shown in Figure 2.

In our scenario, the proposed FL framework decomposes the total electric load into the appliance-wise ones. The usage status of appliances mentioned in the analysis can be visually classified as power-on and power-off status, therefore makes it sensible to characterize the operation of appliances in a simplified way of binary form, thus transforming it into a classification task.

The optimal selection and further advancement of the framework could be done by comparing the true status of appliance-wise with the status derived from the NILM based on the FL framework.

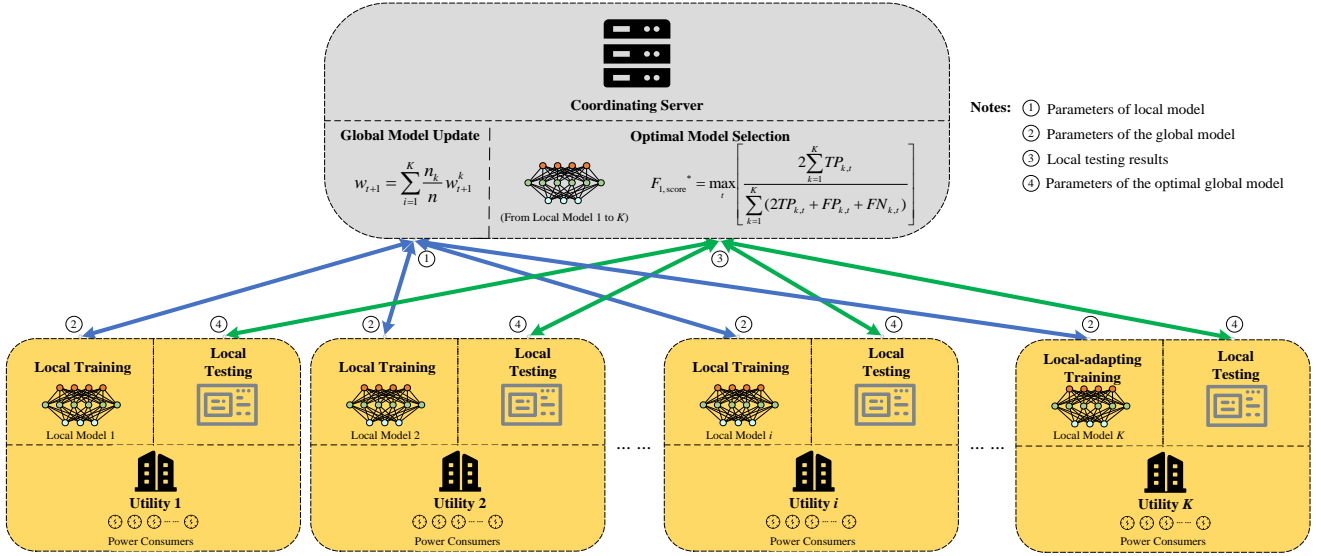

Fig. 2. Structure of the FL framework

### B. Procedure of the Framework

In the framework, there are several local runners that are utilities and DNOs in practical and a coordinating server maintained by a third party. Under the coordination of the server, local networks are trained by all runners using their local NILM datasets. Each target appliance corresponds to a locally-trained network that undergoes supervised training on local examples. Each local network is updated according to the information from the coordinating server.

The procedure of the proposed framework is shown below:

Step 1. Local runners update models based on local electric load datasets.

Step 2. The model parameters are sent to the coordinating server, and the global model is computed using FedAvg. The global model is sent to all local runners.

Step 3. Local runners test their locally-trained models based on local electric load data and send feedbacks to the coordinating server.

Step 4. The coordinating server computes a global testing result based on feedbacks from runners in this epoch. The global testing result computed by the coordinating server reflects the performance of the FL model on the global electric load dataset instead of the local ones.

Step 5. The coordinating server selects the FL model with the best global performance as the optimal model.

The process above repeats until a stopping criterion is met (a maximum number of training iterations).

The procedure of the framework is described in algorithm 1. In the framework, the server is responsible for obtaining the global model through the weighted average gradient, and comprehensively analyzing the global testing of each global model obtained from different rounds. For local runners, not only they need to update their models' parameters through training the local electric load data but also need to test global FL models on their datasets. The coordinating server and local runners communicate with each other to exchange local and global parameters. The details are introduced in subsection C.

### C. Details of the FL Framework

#### 1) Federated Averaging

To make full use of local data from different runners, the proposed framework adopts federated averaging (FedAvg) algorithm.

For local runners, the "Client Update" process is executed in multiple iterations. The $k$-th local runner holds $N_k$ electric load profiles. Local runners split their data into batches in size $B$ and take one step of gradient descent on the current model using local data. $E$ is the number of training epochs each local runner performs over its electric load dataset. Then local runners send updated gradients to the coordinating server.

For the coordinating server, it takes a weighted average of the locally-trained models and sends back model updates to local runners in each round. The global model obtained by FedAvg combines the information of all local electric load datasets. It lays a vital foundation for getting a more generalized model afterwards.

#### 2) Optimal Model Selection

To find the optimal models in multiple rounds, a mechanism that designed to choose optimal model among distributed unseen local data is set up in the proposed framework.

The evaluation of NILM tasks is as follows. To evaluate the performance, the inputs are aggregated electric load data for the testing process of runners. At each time step, a single sample of aggregated electric load data is the input of the network, and the output is the condition of the targeted appliance.

The local runners test the global models received from the coordinating server. The coordinating server computes the global testing result based on the local ones.

As the datasets are not shared among local runners, the local testing result can only reflect the model's performance on local data instead of the global one. It is required to obtain the performance of a locally-trained model on global data, so the global testing result according to the local ones is obliged to acquire. The global testing result can reflect and validate how well the networks generalize to appliances in household unseen in all training data.

In classification tasks, the popular metric F-measure that noted as $F_{1,score}$ can be viewed as the weighted harmonic mean of precision and recall indicator. Precision and recall indicator are usually mutually constrained in massive electric load datasets. An ideal classification performance expects both indicators to be high, but generally, recall is low when precision is high and vice versa. $F_{1,score}$ is the chosen evaluation metric in this paper, it is computed as follows,

$$F_{1,score} = \frac{2 \cdot TP}{2 \cdot TP + FN + FP} \quad (5)$$

where the value of true positive (TP) classifies a positive class as a positive class, the value of true negative (TN) classifies a negative class as a negative class, the value of false positive (FP) classifies a negative class as a positive class, the value of false negative (FN) classifies a positive class as a negative class.

In the proposed framework, the optimal $F_{1,score}$ is noted as follows,

$$F_{1,score}^* = \max_t \left[ \frac{2\sum_{k=1}^{K} TP_{k,t}}{\sum_{k=1}^{K}(2TP_{k,t} + FP_{k,t} + FN_{k,t})} \right] \quad (6)$$

$$t^* = \arg\max_t \left[ \frac{2\sum_{k=1}^{K} TP_{k,t}}{\sum_{k=1}^{K}(2TP_{k,t} + FP_{k,t} + FN_{k,t})} \right] \quad (7)$$

where $t^*$ denotes that in $t$-th round the optimal model is obtained, $k$ denotes the $k$-th utility, $t$ denotes the $t$-th round.

$$w^* = w_{t^*} \quad (8)$$

where $w^*$ denotes parameters of the optimal model, $t$ denotes the $t$-th round.

**Algorithm 1 FedAvg Algorithm**
1: **Server executes:**
2:    initialize $w_0$;
3:    **for** each round $t = 1, 2, \ldots$ **do**
4:       $S \leftarrow$ (set of global model's parameters $w_t$ in this round);
5:       **for** each local runner $k$ **in parallel do**
6:          $w_{t+1}^k \leftarrow$ Client Update($k, w_t$);
7:       **end for**
8:       $w_{t+1} \leftarrow \sum_{k=1}^{K} \frac{n_k}{n} w_{t+1}^k$ ;
9:       **for** each local runner $k$ **in parallel do**
10:       $FN_{k,t}, TP_{k,t}, FP_{k,t} \leftarrow$ Local Testing ($k, t, w_t$)
11:       **end for**
12:    **end for**
12:    **for** each global model's parameters $w_t \in S$ **do**
13:       $w^* \leftarrow$ Optimal Model Selection ($S$)
14:    **end for**
15:
16: **Optimal Model Selection** ($S$): // *Executed on server to get the optimal global model*

$$t^* = \arg\max_t \left[ \frac{2\sum_{k=1}^{K} TP_{k,t}}{\sum_{k=1}^{K}(2TP_{k,t} + FP_{k,t} + FN_{k,t})} \right]$$

17:    $w^* = w_{t^*}$
18:    return $w^*$ to the server
19:
20: **Local runners execute**:
21:    **Client Update** ($k, w$): // *Executed on local runners k*
22:       **for** each local epoch $i$ from 1 to $E$ **do**
23:          batches $\leftarrow$ (data $P_k$ split into batches of size $B$);
24:       **for** batch $b$ in batches **do**
25:          $w \leftarrow w - \eta \cdot \delta l_e$;
26:       **end for**
27:    **end for**
28:    return $w$ to server
29:
30:    **Local Testing** ($k, t$): // *Executed on local runners k to get all the local test result of other models.*
31:    return $FN, TP, FP$

## IV. EXPERIMENT

### A. Data Preparation

The aim is to propose a safe and effective method for utilities and DNOs in data cooperation that could help them accomplish better training based on abundant data.

The scenario is assumed as follows: 4, 8, 16, 32 runners (utilities or DNOs in reality) each holds households' power consumption in their jurisdictions, and they hope to train a preferable model through the corporation in the meantime maintaining their data control and consumers' privacy. This model is expected to be applied to more unseen households. In the experiment, the training data of runners and the testing data are both chosen from REFIT dataset.

REFIT dataset is collected in Loughborough, England, in which containing 20 households' situation from 2013 to 2015 [24]. The active power of the main power supply and single devices is sampled every 8 seconds. This dataset is often used to train deep learning models. same as reference [10], each data sample has 599 dimensions. 1024 continuous electric load profiles from household 3, 4, 5, 7 are chosen for each local runner, and 768000 continuous electric load profiles are chosen evenly from household 3, 4, 5, 7, 19 to set up the testing dataset. It is expected that the model trained based on electric load data from household 3, 4, 5, 7 can generalize well on household 3, 4, 5, 7 and a new household 19. For each appliance, data volume of each local runner is the same. The specific experimental settings are shown in Table II.

### B. Experimental Setting

The experiment runs on the coordinating server with one

2080Ti GPU and seven CPUs. Four scenes are set in which 4, 8, 16, 32 local runners are separately set. And one coordinating service is set. For the coordinating server, FedAvg is carried to coordinate the update of gradients of the locally-trained models. For the runners, NILM models are trained locally. Pytorch is used in the experiment.

The architecture of seq2point learning [10] is used for local runners. A fixed-length window of aggregate active power consumption signal is given as input. The sample window has 599 data points. A sample window was generated by sliding the window forward by a single data point. For the target appliance, the windows of mains are used as inputs of networks, while the midpoints status of the corresponding windows is the output.

### C. Federated Experiments

This subsection illustrates the conduction of the NILM FL experiment. The epoch of the coordinating server is set to be 15. For each local runner, it runs for 10 epochs. The ADAM optimizer algorithm [25] is applied for training. The hyperparameters for training and optimizer parameters are shown in Table I. Eventually, 5 optimal FL models for each appliance are obtained.

TABLE I TRAINING PARAMETERS

| Input window size | 599 |
| --- | --- |
| Number of epochs (local runners) | 10 |
| Batch size | 512 |
| Learning rate of FL | $5\times 10^{-4}$ |
| Number of epochs (coordinating server) | 15 |

To evaluate the performance of these FL models, all of them are compared with the average performance of the locally-trained models and the model trained at the coordinating server with the entire training dataset i.e., the centrally-trained model.

## V. RESULT

As shown in Table II, models trained by FedAvg, centrally-trained model and average performance of locally-trained models are compared in detail. To reveal results more clearly, the improvement rate of $F_{1, score}$ in FedAvg over the centrally-trained and the average performance of locally-trained ones are also presented in Table II.

For each appliance, the centrally-trained models are generally better than the average performance of locally-trained models on $F_{1, score}$ for abundant training data leads to better generalizations. The global model obtained by FedAvg improves the average performance of the locally-trained model to varying degrees as the number of clients increases.

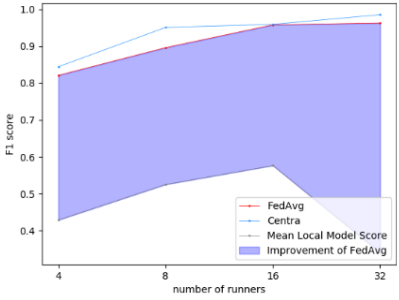
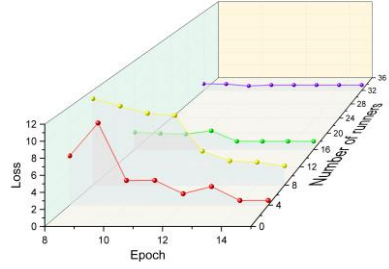
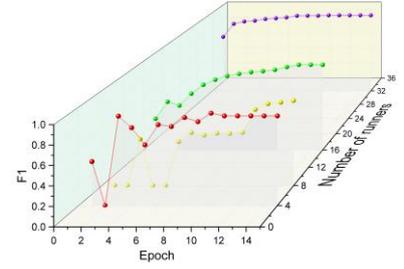

Appliance: Microwave

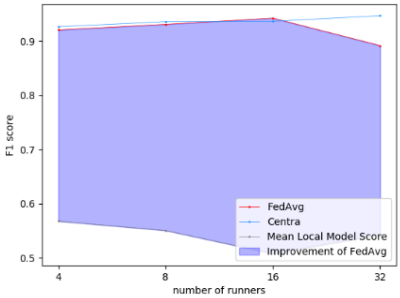
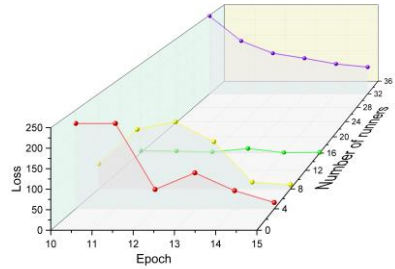
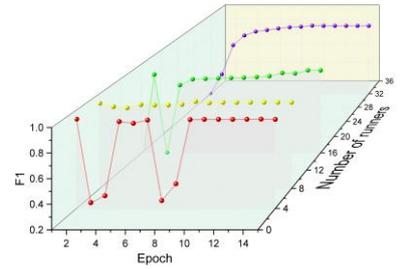

Appliance: Washing machine

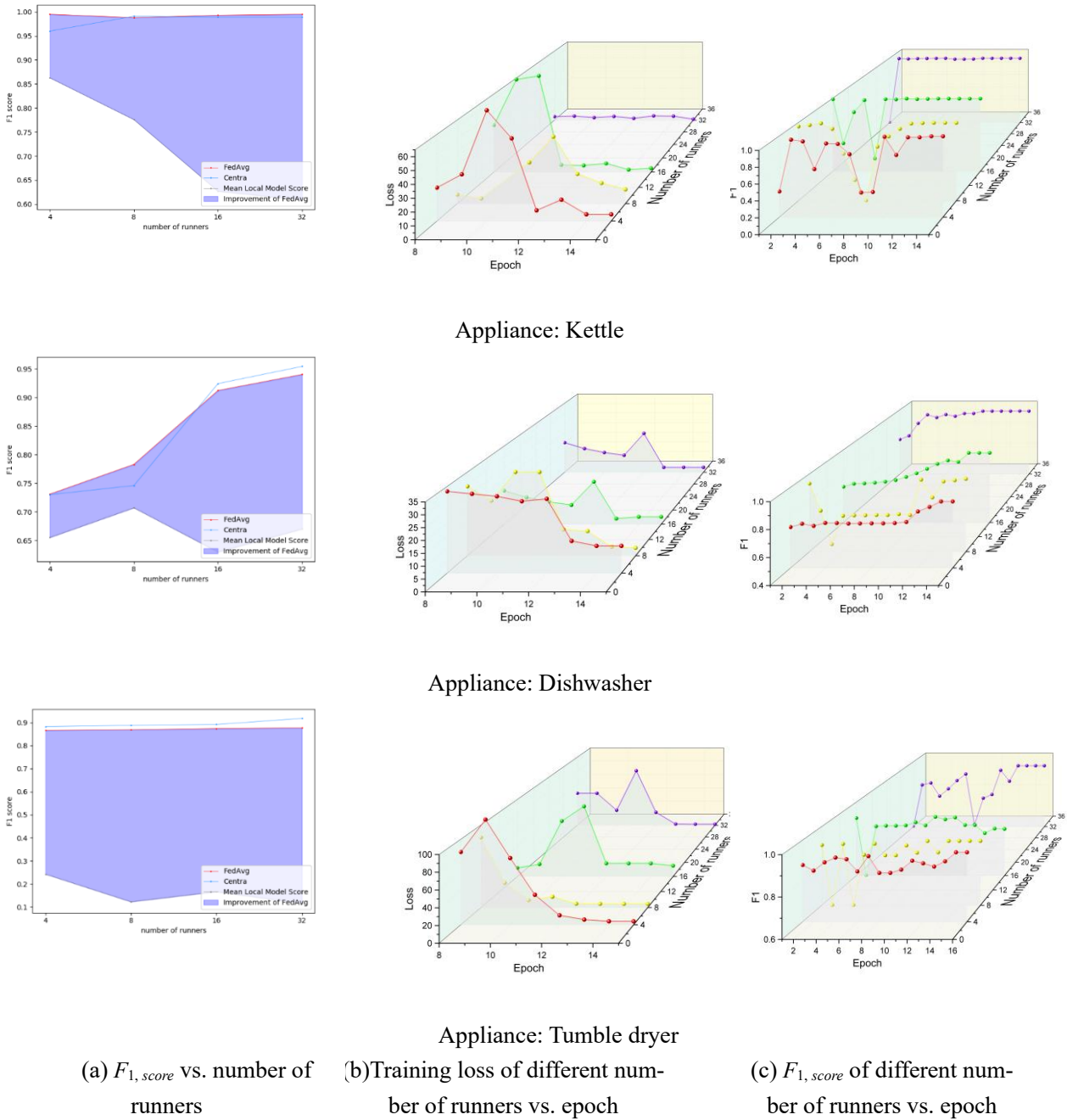

(a) $F_{1,\,score}$ vs. number of runners  (b) Training loss of different number of runners vs. epoch  (c) $F_{1,\,score}$ of different number of runners vs. epoch

Fig. 3. Performances of Appliances modeled by FedAvg, Centrally-trained, and Locally-trained Models

In Figure 3(a), the purple block stands for the improvement of FedAvg over the average performance of locally-trained models. In the activation part of each appliance, regardless of the number of local runners, $F_{1,\,score}$ of FedAvg outclasses the average $F_{1,\,score}$ of locally-trained models, and approximates the centrally-trained model. This indicates that NILM that employees FedAvg has a better performance in identifying and extracting the appliance-wise consumption status form the total electric load reading. It is noticeable that as the number of runners goes up, $F_{1,\,score}$ of all the FedAvg model and centrally-trained model goes up. Especially for appliances such as dishwasher and microwave, $F_{1,\,score}$ of FedAvg and centrally-trained model improves significantly with the increase in the number of runners.

In Figure 3(b), the red, yellow, green, and purple line represent the changes in the training loss value with the training epochs when the number of runners is 4, 8, 16, 32, respectively. Although the training loss value of different numbers of runners fluctuates with the epoch, the overall trend of the loss value decreases along with the increase in epoch, which means that as epochs increase, FedAvg diverge steadily.

In Figure 3(c), the red, yellow, green, and purple line represent the changes in the $F_{1,\,score}$ of FedAvg with the training epochs when the number of runners is 4, 8, 16, 32, respectively. Although the $F_{1,\,score}$ of different numbers of runners fluctuates with the epoch, the overall trend of $F_{1,\,score}$ increases along with the increase in epoch, which means that FedAvg performs better and more steadily.

The comparison of $F_{1,\,score}$ for appliances obtained by different models is shown in Table II.

For the microwave, $F_{1,\,score}$ of FedAvg reaches the top value of 0.963 in the context of 32 local runners. It exceeds by 183.235% compared with the average performance of all the 32 locally-trained models, which demonstrates an outstanding superiority over the locally-trained model, and is 2.333% lower than the $F_{1,\,score}$ of the centrally-trained model.

For the washing machine, $F_{1, score}$ of FedAvg reaches the top value of 0.943 in the context of 16 local runners. It exceeds by 85.897% over the average performance of all the 16 locally-trained models, and is 0.630% higher than the $F_{1, score}$ of the centrally-trained model. This means that FedAvg and centrally-trained model roughly have a similar performance, which reflects the effectiveness of the proposed framework.

For the kettle, $F_{1, score}$ of FedAvg reaches the top value of 0.995 in the context of 4 local runners. It exceeds by 15.255% over the average performance of all the 4 local models, and is 3.608% slightly higher than the $F_{1, score}$ of the centrally-trained model. FedAvg and centrally-trained model almost share a same performance to this extent. When the number of local runners goes up to 8, 16 and 32, the performance is close to that of 4 local runners, which indicates that the data volume of 4 local runners is sufficient for feature extraction of the kettle.

For the dishwasher, $F_{1, score}$ of FedAvg reaches the top value of 0.940 in the context of 32 local runners. It exceeds by 40.299% over the average performance of all the 32 locally-trained models, which is significantly preferable, and it is 1.468% lower than the $F_{1, score}$ of the centrally-trained model, reflecting that the proposed framework helps the training performance approach much close to the centrally-trained model.

For the tumble dryer, $F_{1, score}$ of FedAvg reaches the top value of 0.877 in the context of 32 local runners. It has increased by 662.609% compared with the average performance of all the 32 local models, which is significantly preferable, and it is 4.466% lower than the $F_{1, score}$ of the centrally-trained model, reflecting that the proposed framework helps the training performance approach much close to the centrally-trained model. When the number of runners goes to 4, 8, 16, the performance is close to that of 32 runners, which indicates that the data volume of 4 local runners is sufficient for feature extraction of the tumble dryer.

Considering Figure 3, 4, 5, 6, 7 and Table II, we can find the locally-trained models have different performance due to different types of appliance, but they all perform poorly, requiring the cooperation of multiple local load data. When the data of multiple runners are centrally trained, the model performance is significantly better than the average level of locally-trained models, which proves the necessity of data cooperation. And the performance of FedAvg varies with the number of cooperating local runners and the types of electrical appliances. But they all have greatly improved the average performance level of locally-trained models, which reflects the effectiveness of the proposed framework. The comparison with the centrally-trained model reflects the degree of convergence with the centrally-trained model, thus reflecting how efficient the framework is. At the same time, the experimental results reflect the training situation when the number of local runners is different, which also reflects the diminishing marginal value of the performance improvement of the selected local data in the NILM model training process.

TABLE II COMPARISON OF $F_{1, score}$ OF APPLIANCES OBTAINED BY DIFFERENT MODELS

| Appliance | Number of Runners | Average of Locally-trained Models | Centrally-trained Models | FedAvg | FedAvg over Average of Locally-trained Models (%) | FedAvg over Centrally-trained (%) | Scale of Local Datasets |
|---|---|---|---|---|---|---|---|
| Microwave | 4 | 0.428 | 0.844 | 0.821 | 91.822 | -2.725 | 20480 |
| | 8 | 0.525 | 0.951 | 0.896 | 70.667 | -5.783 | 40960 |
| | 16 | 0.576 | 0.959 | 0.958 | 66.319 | -0.104 | 81920 |
| | 32 | 0.340 | 0.986 | 0.963 | 183.235 | -2.333 | 163840 |
| Washing Machine | 4 | 0.568 | 0.927 | 0.921 | 62.095 | -0.637 | 20480 |
| | 8 | 0.550 | 0.936 | 0.931 | 69.291 | -0.524 | 40960 |
| | 16 | 0.507 | 0.937 | 0.943 | 85.897 | 0.630 | 81920 |
| | 32 | 0.543 | 0.947 | 0.892 | 64.250 | -5.818 | 163840 |
| Kettle | 4 | 0.863 | 0.960 | 0.995 | 15.255 | 3.608 | 20480 |
| | 8 | 0.776 | 0.991 | 0.988 | 27.375 | -0.232 | 40960 |
| | 16 | 0.627 | 0.989 | 0.993 | 58.328 | 0.404 | 81920 |
| | 32 | 0.609 | 0.989 | 0.995 | 63.419 | 0.607 | 163840 |
| Dishwasher | 4 | 0.655 | 0.730 | 0.731 | 11.603 | 0.137 | 20480 |
| | 8 | 0.707 | 0.746 | 0.783 | -3.395 | -8.445 | 40960 |
| | 16 | 0.630 | 0.924 | 0.912 | 44.762 | -1.299 | 81920 |
| | 32 | 0.670 | 0.954 | 0.940 | 40.299 | -1.468 | 163840 |
| Tumble Dryer | 4 | 0.242 | 0.883 | 0.866 | 271.488 | 1.812 | 20480 |
| | 8 | 0.122 | 0.888 | 0.869 | 650.820 | 3.153 | 40960 |
| | 16 | 0.166 | 0.892 | 0.874 | 426.506 | -2.018 | 81920 |
| | 32 | 0.115 | 0.918 | 0.877 | 662.609 | -4.466 | 163840 |

## VI. CONCLUSION

This work establishes a co-modelling mechanism for NILM based on FL. While performing well in NILM, the labelling cost of deep learning based on large training datasets is high. Therefore, data cooperation across runners is crucial to the training process of NILM models. The goal of the proposed framework is to enhance the collaboration between utilities without risks of consumer privacy leakage and loss of data control.

In the framework proposed, the risk of consumer privacy breaches and loss of data control is considerably reduced as local data never leaves the respective utility servers.

In the experiments, the effectiveness of FL in improving the performance of local NILM has been confirmed as well as its ability to handle data from different domains. The model outperforms the locally-trained model, which demonstrates the feasibility of our proposal in enhancing cooperation across utilities. Experimental results show that the results of our framework approximate models trained at the centralized server, demonstrating the efficiency of our framework.

The future work will be focused on combining FedAvg, domain-adaptation for each power consumer, and feedback mechanisms for all local utilities and DNOs. The per-user domain-adaptation will facilitate the adaptation of the global model to the local data thus speeding up the co-modelling process. The feedback mechanism will help the server to find the best-performing locally-trained model, which would lay out the foundation for further use.

In the future, an increasing number of scholars will explore more effective FL methods on communication to improve the performance of NILM models. In the process, the goal will always be the trade-off among privacy protection, model performance and effectiveness. It is also worthwhile to investigate how the global model and locally-trained models after local-adaption can be combined more efficiently.

ACKNOWLEDGMENT

The authors would like to thank…